\documentstyle[proceedings,numreferences]{crckapb}
\newcommand{\barbox}{\stackrel{-}{\Box}}
\begin{opening}
\title{A GEOMETRIC APPROACH TO THE QUANTUM \protect\\
MECHANICS OF DE BROGLIE AND VIGIER}
\author{W.R. WOOD}
\institute{Faculty of Natural and Applied Sciences\\ 
Trinity Western University, 7600 Glover Road\\ 
Langley, British Columbia V2Y 1Y1, Canada}
\author{G. PAPINI}
\institute{Department of Physics, University of Regina\\
Regina, Saskatchewan S4S 0A2, Canada}
\end{opening}

\runningtitle{A GEOMETRIC APPROACH}
\begin{document}

\begin{abstract}
Following de Broglie and Vigier, a fully relativistic
causal interpretation of quantum mechanics is given
within the context of a geometric theory of gravitation and electromagnetism.  While
the geometric model shares the essential principles of the causal interpretation
initiated by de Broglie and advanced by Vigier, the particle and wave components
of the theory are derived from the Einstein equations rather than a nonlinear wave 
equation.  This geometric approach leads to several new features, including a 
solution to the de Broglie variable mass problem.
\end{abstract}

\section{Introduction}
It is a pleasure to acknowledge the role that
Professor Vigier has played in the development of the casual interpretation of
quantum mechanics \cite{Hol93}.  His demonstration of an 
explicit solitonic solution \cite{Vig91} has
made de Broglie's conception of a double 
solution \cite{Vigier} a reality.  As well, his extensive
work \cite{Ann} on issues relating to relativistic causal or stochastic models has been very 
helpful in our own efforts to formulate the principles of the causal 
interpretation within a geometrical framework.

In the geometric theory discussed here,
a particle is represented by a thin shell or bubble solution to the 
Einstein equations rather than a solitonic solution to a nonlinear
wave equation.    
The Gauss-Mainardi-Codazzi (GMC) formalism (a familiar
tool in general relativity) is used to 
facilitate the analysis of the dynamics of the bubble.
The junction conditions in the GMC formalism provide
a tensorial description of the balance of energy and momentum
across the thin shell.  As a consequence, the geometric model 
provides a framework by which the influence of external fields, 
such as the wave field $\psi (x)$, on the motion of the
particle can be rigorously analyzed. 

In the classical theory of general relativity, the guidance 
mechanism is well-known:  the geometry, which is determined by
the distribution of matter, in turn, governs the motion of the matter
itself.  However, this classical guidance mechanism becomes 
insignificant when applied to particles at the microscopic scale
where de Broglie's guidance principle is required to explain
quantum effects such as the interference pattern in the two-slit
experiment.  It appears that a theory of gravitation whose domain
of validity encompasses the microscopic scale is required if the
desired guidance mechanism is to be given a geometric interpretation.
A natural candidate is Weyl's conformally invariant theory of
gravitation and electromagnetism \cite{Wey18}.  
Weyl generalized the Riemannian
geometry of general relativity by supposing that a vector parallel
transported around a closed circuit would also experience a change
in length according to the formula $\delta \ell =\ell \kappa _{\mu }
\delta x^{\mu }$.  The vector field $\kappa ^{\mu }$, together with
the metric tensor $g_{\mu \nu}$ that is defined modulo an
equivalence class, comprise the fundamental fields of the new
geometry.  The choice of Weyl geometry is also strongly supported
by Santamato's demonstration \cite{San84} 
of a (nonrelativistic) ``quantum force'' associated
with $\kappa _{\mu }$ in his stochastic theory of ``geometric
quantum mechanics'' \cite{San85}.  In the model presented here, the 
geometry of Weyl is used to express the principles of the causal interpretation
of quantum mechanics in a fully relativistic form.

Apart from providing a means to investigate the self-consistency of
the dynamical aspects of the causal interpretation, the geometric 
model also offers several new interesting features.  For example, 
by formulating the theory in the context of curved spacetime, new
opportunities arise for considering the role that nonlocal interactions
may play in a relativistic causal theory.  As well, the geometric model
provides a resolution to the problem of de Broglie's variable mass.

\section{The GMC Formalism}
In Weyl geometry, one introduces a gauge-covariant 
calculus \cite{Gre} based on the
gauge-covariant derivative $\stackrel{-}{\Box}$
and a semimetric connection 
$\bar{\Gamma }^{\alpha }_{\ \mu \nu }$,
where an overbar is used to distinguish objects from their Riemannian
counterparts.  In the GMC formalism, a timelike
hypersurface $\Sigma $, which represents the history of the thin
shell, divides spacetime into two four-dimensional regions 
($V^{I}$ and $V^{E}$), both of which have $\Sigma $ as their boundary.
The intrinsic metric on $\Sigma $ is given by
$h_{\mu \nu }=g_{\mu \nu }-n_{\mu }n_{\nu }$,
where $n^{\mu }$ is a unit spacelike 
($n_{\mu }n^{\mu }=1$) vector field normal to $\Sigma $.   The
extrinsic curvature tensor in Weyl geometry
is defined by
$\bar{K}_{\mu \nu }=
K_{\mu \nu }+h_{\mu \nu }n^{\alpha }\kappa _{\alpha }$.
The development of the GMC formalism in Weyl geometry ultimately
yields the equations \cite{Woo92}
\begin{eqnarray}
n_{\mu }n^{\nu }G^{\mu }_{\ \nu }=
-\frac{1}{2}(^{3}\hspace{-2pt}R+K_{\mu \nu }K^{\mu \nu }-K^{2})
-D_{\mu }\kappa ^{\mu }+2h_{\mu }^{\ \nu }\kappa ^{\mu }\kappa _{\nu }
+2Kn^{\mu }\kappa _{\mu }, \label{1}
\end{eqnarray}
\begin{eqnarray}
n_{\mu }h_{\alpha }^{\ \nu }G^{\mu }_{\ \nu }=
D_{\alpha }K-D_{\mu }K^{\mu }_{\ \alpha }, \label{2}
\end{eqnarray}
\begin{eqnarray}
h^{\alpha }_{\ \mu }h_{\beta }^{\ \nu }G^{\mu }_{\ \nu }=
\hspace{1pt}^{3}\hspace{-1pt}G^{\alpha }_{\ \beta}&+&
(K^{\alpha }_{\ \beta }-h^{\alpha }_{\ \beta }K)
_{,n}-KK^{\alpha }_{\ \beta }+\frac{1}{2}h^{\alpha }_{\ \beta }
(K_{\mu \nu }K^{\mu \nu }+K^{2}) \nonumber \\
&-&2(K^{\alpha }_{\ \beta }-h^{\alpha }_{\ \beta }K)n^{\lambda }
\kappa _{\lambda }+2h^{\alpha }_{\ \beta }h_{\mu }^{\ \nu }
\kappa ^{\mu }\kappa _{\nu }. \label{3}
\end{eqnarray}

The intrinsic stress-energy tensor on $\Sigma $, which is defined by
\begin{eqnarray}
S^{\mu }_{\ \nu }\equiv \lim _{\varepsilon \rightarrow 0}
\int _{-\varepsilon }^{\varepsilon }T^{\mu }_{\ \nu }dn, \label{4}
\end{eqnarray}
corresponds to the distributional part of $T_{\mu \nu }$.
The junction conditions for the gravitational field are given by
$h^{\alpha }_{\ \mu }h_{\beta }^{\ \nu }S^{\mu }_{\ \nu }=
\gamma ^{\alpha }_{\ \beta }-h^{\alpha }_{\ \beta }\gamma  $ and 
$n_{\mu }S^{\mu}_{\ \nu }=0$, 
where the jump in the extrinsic curvature is denoted by
$\gamma ^{\mu }_{\ \nu }\equiv [K^{\mu }_{\ \nu }]$,
$\gamma \equiv \gamma ^{\mu }_{\ \mu }$, and $g_{\mu \nu }$ and
$\kappa ^{\mu }$ are assumed to be 
continuous across $\Sigma $, but their
normal derivatives discontinuous.  It is also assumed that 
$\kappa _{\mu }=0$ in the interior geometry $V^{I}$ so that length
integrability is established in the spacetime region occupied by the
particle.  Using (\ref{1}) and (\ref{2}), the jump in the equations
$n_{\mu }G^{\mu }_{\ \nu }=n_{\mu }T^{\mu }_{\ \nu }$ yields
the intrinsic tensor equations 
\begin{eqnarray}
D_{\mu }(h^{\mu }_{\ \alpha }h_{\nu }^{\ \beta }
S^{\alpha }_{\ \beta })+[n_{\alpha }h_{\nu }^{\ \beta }
T^{\alpha }_{\ \beta }]=0, \label{5}
\end{eqnarray}
\begin{eqnarray}
\{K^{\mu }_{\ \nu }\}S_{\mu }^{\ \nu }+
[n_{\mu }n^{\nu }T^{\mu }_{\ \nu }]=0, \label{6}
\end{eqnarray}
where $\{ K^{\mu }_{\ \nu } \}$ denotes the average of 
$K^{\mu }_{\ \nu }$ across $\Sigma $.
Equations (\ref{5}) and (\ref{6}) describe the balance of
stress-energy-momentum between neighboring external fields and the
thin shell.  It is this balance that governs the dynamical
behavior of the thin shell.  Indeed, the requirement that
the fields in the exterior Weyl space join at $\Sigma $
in accordance with the junction conditions places constraints
on the motion of the bubble since $\Sigma $ represents 
the {\em history} of the thin shell.  
Within the context of the causal interpretation of quantum mechanics,
it is particularly significant that the interplay between the
particle and wave aspects of the problem is an inherent
feature of the present geometric formulation.

\section{The Geometric Model}
Once the assumption is made that elementary particles follow trajectories that
are influenced in part by a wave field $\psi (x)$, it is only natural to consider the new
field on equal footing with the gravitational and electromagnetic fields in a unified
manner.  In fact, the transfer of energy and momentum required
in the guidance process suggests use of a tensorial formulation that would, hopefully, yield the Einstein-Maxwell theory in the classical limit.  The fact that the hypothesized guidance mechanism is effective at the microscopic scale, while the corresponding mechanism in general relativity is significant only at large scales, suggests beginning with a conformally invariant theory to integrate quantum effects into a geometric theory.  In this regard, it is of interest to note that under the local conformal transformation 
$g_{\mu \nu }\rightarrow \rho ^{2}g_{\mu \nu }$, the scalar curvature transforms as $R\rightarrow R+\frac{6}{\rho}\Box _{\mu }\Box ^{\mu }\rho $, where the derivative term in $\rho $ is the covariant generalization of the quantum potential in the
causal interpretation of quantum mechanics.  The conformally invariant geometry
introduced by Weyl is particularly attractive because it also provides a geometric
interpretation for the electromagnetic field.

For our purposes, the modified Weyl-Dirac theory \cite{Gre}
\begin{eqnarray}
I_{c}=\int \left\{ -\frac{1}{4}f_{\mu \nu }f^{\mu \nu }\hspace{-2pt}+
|\beta |^{2}\bar{R}+ 
k|\hspace{-2pt}\stackrel{-}{\Box} _{\mu }\hspace{-2pt}\beta \stackrel{-}{\Box} 
\rule{0pt}{0pt}\hspace{-2pt}^{\mu }\beta|+
\lambda|\beta |^{4} \right. \nonumber \\
\hspace*{1 cm}\left. + \rho \gamma ^{\mu }
(\stackrel{-}{\Box}_{\mu }\hspace{-2pt}\rho - \varepsilon \rho \varphi _{,\mu })
\right\} \sqrt{-g}d^{4}x, \label{7}
\end{eqnarray}
where $\varepsilon =\pm 1$, $k$ and $\lambda $ are real arbitrary
constants and $\rho $, $\varphi $ and $\kappa _{\mu }$ are real fields,
is convenient because it gives the complex scalar field $\beta =\rho e^{i\varphi }$ a geometrical status as well as maintaining a theory
that is linear in the scalar curvature.  This latter point is essential when the particle is
associated with a region of Riemannian space 
where the conformal symmetry of the exterior Weyl space is broken and the 
Gauss-Mainardi-Codazzi (GMC) formalism \cite{Woo92} is used to join the interior and exterior regions. The constraint,
$\kappa _{\mu }=-(\ln \rho )_{,\mu }+\varepsilon \varphi _{,\mu } $,
is introduced \cite{Gre} to allow for quantization of flux and leads to a topologically
nontrivial electrodynamics with $\varepsilon f_{\mu \nu }=
\varphi _{,\nu \mu }-\varphi _{,\mu \nu }$. 

\subsection{THE FIELD EQUATIONS}
The field equations that follow
from the action (\ref{7}), given here in terms of the Riemannian fields, are \cite{Gre}
\begin{eqnarray}
\Box _{\nu }f^{\mu \nu }=
4(k-3\varepsilon )\rho ^{2}\varphi ^{,\mu }
\equiv j^{\mu }, \label{8}
\end{eqnarray}
\begin{eqnarray}
G_{\mu \nu }=\frac{1}{2\rho ^{2}}E_{\mu \nu }+
I_{\mu \nu }+\frac{1}{2}\lambda g_{\mu \nu }\rho ^{2}+
H_{\mu \nu }\equiv T_{\mu \nu }, \label{9}
\end{eqnarray}
\begin{eqnarray}
\frac{1}{3}(\varepsilon k-3)\varphi _{,\mu }\varphi ^{,\mu }=
-\frac{1}{6}(R+2\lambda \rho ^{2})+
\frac{1}{\rho }\Box _{\mu }\Box ^{\mu }\rho \label{10}
\end{eqnarray}
and
\begin{eqnarray}
(k-3\varepsilon )\Box _{\mu }(\rho ^{2}\varphi ^{,\mu })=0, \label{11}
\end{eqnarray}
where $E_{\mu \nu }$ is the usual Maxwell tensor, 
\begin{eqnarray}
I_{\mu \nu }=\frac{2}{\rho}(\Box _{\nu }\Box _{\mu }\rho -
g_{\mu \nu }\Box _{\alpha }\Box ^{\alpha }\rho )-
\frac{1}{\rho ^{2}}(4\rho _{,\mu }\rho _{,\nu }-
g_{\mu \nu }\rho _{,\alpha }\rho ^{,\alpha }) \label{12}
\end{eqnarray}
and
\begin{eqnarray}
H_{\mu \nu }=-2(\varepsilon k-3)
(\varphi _{,\mu }\varphi _{,\nu }-
\frac{1}{2}g_{\mu \nu }\varphi _{,\alpha }\varphi ^{,\alpha }). \label{13}
\end{eqnarray}
Taking the trace of (\ref{9}) one recovers (\ref{10}), while
(\ref{11}) follows from the conservation equation associated with (\ref{8}).
The theory also contains the wave equation \cite{Pap89}
\begin{eqnarray}
(\Box ^{\lambda }+i\kappa ^{\lambda })
(\Box _{\lambda }+i\kappa _{\lambda })\psi -
\frac{\lambda }{3}|\psi |^{2}\psi -
\frac{1}{6}R\psi =0. \label{14}
\end{eqnarray}
In fact, if one writes $\psi =\rho e^{i\chi }$ and defines
$\chi _{,\mu }$ according to $\alpha \varphi _{,\mu }\equiv
\chi _{,\mu }+\kappa _{\mu }$ with $\alpha ^{2}\equiv
(\varepsilon k-3)/3$, then
the imaginary part of (\ref{14}) yields (\ref{11}),
while the real part coincides with (\ref{10}) which can be expressed as
\begin{eqnarray}
(\chi _{,\mu }+\kappa _{\mu })
(\chi ^{,\mu }+\kappa ^{\mu })=
-\frac{1}{6}(R+2\lambda \rho ^{2})+
\frac{1}{\rho }\Box _{\mu }\Box ^{\mu }\rho .\label{15}
\end{eqnarray}
Since $\varphi $, $\chi $, and $\kappa _{\mu }$
are real fields, $\alpha $ must be a real constant.
In the causal interpretation of quantum mechanics, equation (\ref{15})
is identified as the Hamilton-Jacobi equation for a system of momentum
$\chi ^{,\mu }+\kappa ^{\mu }=Mu^{\mu }$, so that
\begin{eqnarray}
Mu_{\mu }=\alpha \varphi _{,\mu }. \label{16}
\end{eqnarray}
From (\ref{15}) one finds
\begin{eqnarray}
M^{2}=\frac{\lambda }{3}\rho ^{2}+
\left( \frac{R}{6}-\frac{1}{\rho }\Box _{\mu }
\Box ^{\mu }\rho \right)  , \label{17}
\end{eqnarray}
which is the square of the de Broglie mass in the present model.

\subsection{THE PARTICLE-WAVE SOLUTION}  
For $\alpha \varphi _{,\mu }=Mu_{\mu }$, the tensor $H_{\mu \nu }$ 
is seen to represent a perfect (irrotational)
fluid with equal pressure and 
energy density:  $H_{\mu \nu }=-6M^{2}
(u_{\mu }u_{\nu }+\frac{1}{2}g_{\mu \nu })$. 
In the present geometric model, 
$H_{\mu \nu }$ is identified with the 
Madelung fluid in the causal interpretation. 
The particle is represented by a static, spherically symmetric
thin shell solution to the Einstein equations when
the Madelung fluid tensor $H_{\mu \nu }$ is neglected.

Application of the GMC 
formalism requires the determination of the
interior and exterior line elements
\begin{eqnarray}
ds^{2}_{I,E}=-e^{\nu _{I,E}}dt^{2}_{I,E}+e^{\mu _{I,E}}dr^{2}+
r^{2}(d\theta ^{2}+\sin ^{2}\theta d\phi ^{2}), \label{18}
\end{eqnarray}
as well as the intrinsic stress-energy tensor $S_{\mu \nu }$
on the timelike hypersurface $\Sigma $.  
In the interior space it is assumed that
$\kappa _{\mu }=0$ and that the scalar field acquires a 
constant value $\rho = \rho _{0}$ which breaks the interior 
conformal invariance and fixes the scale of the particle.  Under
these conditions, the interior metric is given by \cite{Woo92}
\begin{eqnarray}
e^{-\mu _{I}}=1+\frac{1}{6}\lambda \rho _{0}^{2}r^{2}=
e^{\nu _{I}}, \label{19}
\end{eqnarray}
so that the interior space is de Sitter ($\lambda <0$), 
Minkowski ($\lambda =0$) or anti-de Sitter ($\lambda >0$).
The exterior metric, expressed in terms of the arbitrary function
$\rho (r)$, is given by \cite{Woo92}
\begin{eqnarray}
e^{-\mu _{E}}=\left( 1+r\frac{\rho '}{\rho }\right) ^{-2}
\left[ 1-\frac{2m}{\rho r}+\frac{q^{2}}{4\rho ^{2}r^{2}}+
\frac{1}{6}\lambda \rho ^{2}r^{2}\right]  \label{20}
\end{eqnarray}
and 
\begin{eqnarray}
e^{\nu _{E}}=(\ell _{0}\rho )^{-2}
\left[ 1-\frac{2m}{\rho r}+\frac{q^{2}}{4\rho ^{2}r^{2}}+
\frac{1}{6}\lambda \rho ^{2}r^{2}\right] , \label{21}
\end{eqnarray}
where $m$, $q$ and $\ell _{0}$ are integration constants
and a prime denotes differentiation with respect to $r$.
When it is
assumed that $g_{\mu \nu }$ and
$\kappa ^{\mu }$ are continuous across $\Sigma $, but their
normal derivatives discontinuous,  
the surface stress-energy tensor is found to take the form \cite{Woo92}
\begin{eqnarray}
h^{\alpha }_{\ \mu }h_{\beta }^{\ \nu }
S^{\mu }_{\ \nu }=-2\sigma h^{\alpha }_{\ \beta }, \label{22}
\end{eqnarray}
where $\sigma \equiv [n^{\mu }(\ln \rho )_{,\mu }]$; that is,
the surface stress-energy tensor is induced when the normal
derivative of $\ln \rho $ across $\Sigma $ is discontinuous.
The intrinsic stress-energy
tensor (\ref{22}) is characteristic of a domain wall of surface
energy $2\sigma $, where $h_{\mu \nu }$ is the intrinsic metric
on $\Sigma $.  For $\sigma >0$, the bubble is under a surface tension
that opposes the Coulomb repulsion due to the surface
charge.  In this way, the particle finds its origin in the 
field $\rho $ that (i) fixes the scale in $V^{I}$, (ii) ensures
conformal invariance in $V^{E}$, and (iii) induces the surface
tension needed for stability.  

Taking $\varphi _{,\mu }=0$ in $V^{I}$ and 
$h_{\mu }^{\ \nu }\varphi _{,\nu }$ discontinuous across
$\Sigma $ allows the bubble to be embedded
in the Madelung fluid in accordance with (\ref{5}) and (\ref{6}),
while the surface stress-energy tensor (\ref{22}) remains
unchanged.  For $n^{\mu }\varphi _{,\mu }=0$, the normal
component of the exterior fluid momentum at $\Sigma $ takes
the form $n_{\mu }H^{\mu }_{\ \nu }=-3M^{2}n_{\nu }$.  From
this it follows that $n_{\mu }h_{\alpha }^{\ \nu }
H^{\mu }_{\ \nu }=0$ and $n_{\mu }n^{\nu }H^{\mu }_{\ \nu }=
-3M^{2}$.  As a consequence, the Madelung fluid tensor 
$H_{\mu \nu }$ does not contribute to (\ref{5}), the time
component of which, in the rest frame of the thin shell,
governs the transfer of energy between the particle and its
neighboring fields \cite{Bar91}.
While the particle does not
draw energy from $H_{\mu \nu }$ as one might expect \cite{Vig91},
it can be shown that, if $[\rho _{,n}]$ varies on $\Sigma $,
then $I_{\mu \nu }$ will transfer energy to the thin shell
such that $\sigma \neq $ constant.

\subsection{THE GUIDANCE CONDITION}
Although the Madelung
fluid doesn't serve as an energy source for the particle,
it does influence the motion of the thin shell through
(\ref{6}) which represents Newton's second law \cite{Bar91}.
In this manner, the bubble acquires a new
dynamical nature as it is guided in its 
motion by the fluid in $V^{E}$.
The realization of this guidance process can be seen by 
considering the dynamical behavior of a fluid element 
at a point $P$ on the exterior surface of the bubble,
where the four-velocity of the fluid element is 
denoted $u^{\mu }_{f}(P)=dz^{\mu }(P)/ds_{E}$.
By construction, the metric tensor is continuous across
$\Sigma $ at $P$ and consequently,
$ds_{E}^{2}(P)=ds_{\Sigma }^{2}(P)$.  Hence,
at any point $P$ on $\Sigma $, the four-velocity of the
thin shell is given by
\begin{eqnarray}
u^{\mu }_{p}(P)=\left. \frac{dz^{\mu }}{ds_{\Sigma }}
\right| _{P}=\left.
\frac{dz^{\mu }}{ds_{E}}\right| _{P}=u^{\mu }_{f}(P)=
\frac{\alpha }{M}\varphi ^{,\mu }(P) \label{23}
\end{eqnarray}
which is recognized as the guidance formula advanced by
de Broglie.  In the present approach, the validity of the
guidance condition can be extended beyond holding only at 
a given point by noting that the motion of the fluid along its 
worldline from $P$ to a subsequent point $P'$, at which
the fluid element and the bubble are still in contact, can
be viewed as the result of a conformal transformation 
induced by the factor \cite{Pap92}
\begin{eqnarray}
\xi ^{2}=1-M^{2}\left( \frac{\barbox M}{ds}\right) ^{-2}
\left( \frac{\barbox u_{\mu }}{ds}\right) ^{2}. \label{24}
\end{eqnarray}
The metric tensor at $P'$ in $V^{E}$ can therefore be
obtained from its corresponding value at $P$ by applying
a conformal transformation with the same factor $\xi ^{2}$
and, by continuity, the intrinsic metric at $P'$ is also
determined.  The resulting identity, 
$h_{\mu \nu } (P')=\xi ^{2}h_{\mu \nu } (P)$, leads
to the conclusion that the bubble and fluid must move
in step.  Consequently, the Hamilton-Jacobi equation
(\ref{15}) may be applied to the particle itself, as
required in the causal interpretation of quantum mechanics.

\subsection{NONLOCAL EFFECTS AND CURVED SPACETIME}
A novel feature of the bubble model presented above
is the manner in which the interior space $V^{I}$ is made distinct from the
exterior space $V^{E}$.  This property not only makes it possible to break
the conformal invariance in the interior space, whereby standards
of length can be introduced into the theory while Weyl's geometric
interpretation of the exterior electromagnetic field is preserved,
but {\em the locality requirements of the exterior space need not 
be imposed in the interior space}.  In particular, 
nonlocal influences that have 
been observed\footnote{It is
interesting to note, however, that Squires\cite{Squ} has challenged the conclusion 
that the empirical evidence implies nonlocality
by considering the time involved in the actual measuring process in an Aspect-like experiment.}
in experiments employing
correlated particles may simply be
a consequence of the fact that, while the world tubes of the
correlated particles diverge after the disintegration process,
they actually share a common past geometry that affords
nonlocal interactions.  In this
way, nonlocal effects could be explained without denying
the objective reality of elementary particles or compromising
the principles of relativity (in curved spacetime).

The suggestion that the separation of $V^{I}$ and $V^{E}$
plays an essential role in seeking to understand the
intriguing nonlocal EPR-type correlations does not require any
nonlocal effects to occur in the 
exterior Weyl space.  This situation is clearly
not in keeping with the idea 
that it is the quantum potential (that exists in $V^{E}$ in the present model) 
that is responsible for nonlocal phenomena.  Bohm {\em et al \/} \cite{Boh87}
have argued that the invariance of the quantum potential under a
scaling of $\psi (x)$ by an arbitrary {\em constant \/} plays a
fundamental role in the nonlocal nature of the theory.  However,
this invariance property is reminiscent of the global phase
invariance of pre-gauge field theories that also
``contradicts the letter and spirit of relativity'' \cite{Ryd85}, and as a
consequence is replaced by local phase invariance.  In the present geometric
model, the generalized quantum potential in (\ref{17}) is invariant
under the conformal transformation
\begin{eqnarray}
\tilde{g}_{\mu \nu }=\sigma ^{2}g_{\mu \nu },
\quad \tilde{\rho }=\sigma ^{-1}\rho \label{25}
\end{eqnarray}
for the arbitrary {\em function} $\sigma ^{2}(x)> 0$.  
As mentioned earlier, by requiring invariance under
the local scaling (\ref{25}) one is naturally
led to a conformally invariant theory.  For the Weyl-Dirac theory
considered above, information regarding the particle's environment is
propagated in the exterior spacetime
via the tensor field $I_{\mu \nu }$ in a local manner.

\subsection{A SOLUTION TO DE BROGLIE'S VARIABLE MASS PROBLEM}
An outstanding issue in the de Broglie-Vigier causal interpretation of quantum
mechanics \cite{Vigier} has been the problem associated 
with the reality of the variable mass $M$.
Within the context of second-order wave equations, this problem manifests itself
in the mathematical existence of negative probability densities and negative 
energy solutions.  When the usual probabilistic interpretation is applied, these 
solutions cannot, in general, be given a physically meaningful interpretation.
In contrast, when a particle follows a timelike causal trajectory, the situation 
changes radically.  In this case, positive energy solutions are necessarily correlated 
with positive values of $M$ and positive probability densities and the sign
of the energy remains fixed along the trajectory \cite{Dewdney}.
However, general solutions of the Klein-Gordon equation do not ensure the 
reality of $M$.  

This deficiency  is overcome in 
the present geometric model \cite{Woo95} due to the existence of 
the timelike thin shell solution to Einstein's
equations which can be embedded in the Madelung fluid 
according to the junction
conditions discussed above. 
While spacelike and timelike directions are distinguished in 
any relativistic theory, a geometric theory permits these directions 
in spacetime to be related to the motion of matter via 
Einstein's field equations.  That is, for a given foliation of spacetime,
the GMC formalism requires the various timelike and spacelike
components of $G_{\mu \nu }$ to be equated to the corresponding 
components of $T_{\mu \nu }$.  In this way, a link is established
between the properties of spacetime and matter that allows one to
address the issue of whether or not a given four-vector that is associated
with matter is timelike.
It is due to the absence of this geometric structure 
that the possibility of spacelike four-momenta in de Broglie's guidance formula 
$P_{\mu }=Mu_{\mu }$ cannot be excluded in previous
formulations of the causal interpretation derived from a scalar wave equation.
By basing the theory on the field equations (\ref{8})-(\ref{11}) (from which 
the wave equation (\ref{14}) is then identified), one is not bound to demonstrate that all possible generic solutions to the wave equation must be physically meaningful as 
is the case when the causal
interpretation is based solely on a wave equation.  It is the field equations 
(\ref{8})-(\ref{11}) that determine the set of physically acceptable solutions
in the geometric approach.
In the geometric model discussed above, the timelike nature
of $Mu_{\mu }$ can be demonstrated as follows.

The constraint in (\ref{1}) can be written in the gauge 
$\kappa _{\mu }'=\kappa _{\mu }+(\ln \rho )_{,\mu }$
so that $\kappa _{\mu }'=\varepsilon \varphi _{,\mu }$.  
This is permissible due
to the gauge covariance of the theory and this particular gauge is viable
since $\rho $ must be greater than one in the model in order for the thin shell to be under 
a surface tension that balances the Coulomb repulsion \cite{Woo92}.  For the 
static solution of Section 3b, 
$\kappa _{\mu }' \kappa ^{\mu }\hspace{1pt}'<0$ and
as a consequence $\varphi _{,\mu }$ is timelike.    It then follows from (\ref{16}) that $M$ must be real since $\alpha $ is real.  This result, obtained in the static case, also holds
true in a frame comoving with the thin shell, and is therefore quite general.  Indeed, due to the covariant nature of $M$ under conformal tranformations that preserve the sign of the line element, $M^{2}$ must be positive in general.

The condition for $\varphi _{,\mu }$ to be timelike can also be expressed within the 
context of the theorem of Frobenius.  In terms of differential forms, equation (\ref{16})
is given by 
\begin{eqnarray}
u=\frac{\alpha }{M}\hspace{4pt}d\varphi \equiv h\hspace{4pt}d\varphi . \label{26}
\end{eqnarray}
The condition for $u^{\mu }$ to be orthogonal to hypersurfaces of constant $\varphi $, 
and hence for $\varphi _{,\mu }$ to be timelike, is given by
$du \wedge u=0$.  Recognizing that, due to the multivalued nature of $\varphi $, 
$d\varphi $ is not closed even though it is an exact 1-form, the condition
for timelike $\varphi _{,\mu }$ becomes
\begin{eqnarray}
d\hspace{2pt}^{2}\varphi \wedge d\varphi =0.  \label{27}
\end{eqnarray}
Equation (\ref{27}) is satisfied in the static case considered above, where 
$d\hspace{2pt}^{2}\varphi \sim dx^{0}\wedge dx^{1}$ and
$d\varphi \sim dx^{0}$.

For timelike $\varphi _{,\mu }$, it follows that the Maxwell current $j^{\mu }$ in (\ref{8}) is also timelike without having to impose this as an auxiliary condition.  In the present theory, the Maxwell current is proportional to the Klein-Gordon current \cite{Pap89} associated with the wave equation,
$j^{\mu }_{KG}=\alpha \rho ^{2} \varphi ^{,\mu } $,
and is therefore also timelike and as such does
not suffer from the difficulties normally associated with the current for a 
second order wave equation.  It should be noted, however, that $\psi $ in the
present theory is a physical field and not immediately identifiable with a 
probabilistic wave function.
In addition, the time component of the current can be
made positive by choosing the positive sign of the radical in the definition of $\alpha $.  It then follows that positive (negative) energy particles will correspond to positive (negative) values of $M$ and positive (negative) values of $j^{\mu}_{KG}$.  
In this regard, it is interesting to observe that the equation of motion
\begin{eqnarray}
\frac{\barbox }{ds}(Mu_{\mu })=
-\varepsilon \alpha f_{\mu \nu }u^{\nu }-
\barbox _{\mu }M \label{28}
\end{eqnarray}
is invariant under charge conjugation and 
time reversal transformations as discussed by Dirac \cite{Dirac}.  In addition,
equation (\ref{28}) is invariant under $M\rightarrow -M$ together with time reversal.
This indicates that, in the present theory, negative energy particles may be interpreted
as positive energy particles moving backward in time.

\section{Summary}
Although the geometric formulation of the causal theory presented above
is in an early developmental stage, it nevertheless demonstrates that it is possible
to inject the principles of the causal 
interpretation of quantum mechanics into a fully relativistic geometric
theory in Weyl space.  In the authors' opinion, 
this is an essential step towards
obtaining a satisfactory causal theory of 
quantum phenomena.  By formulating
the problem within the context of a theory of gravitation, whereby
the description of the transfer of energy-momentum becomes an 
inherent feature, it becomes possible to demonstrate that
the guidance principle
is dictated by the physics rather than the physicist.
As well, the geometric model provides a basis upon which
issues such as the reality of the de Broglie variable mass and
nonlocal interactions can be addressed.

\vspace{.3cm}
\begin{flushleft}
{\large
{\bf Acknowledgements}}\vspace{.3cm}\\
\end{flushleft}
This work was supported in part by the Natural Sciences and
Engineering Research Council of Canada.  The authors are grateful for
the assistance provided by S.\ Jeffers for the International Organizing
Committee.  One of the authors (G.P.) wishes to thank Dr.\ K.\ Denford,
Dean of Science, University of Regina, for continued research support.

\end{document}